\documentclass[preprint,12pt,authoryear]{elsarticle}

\usepackage{amsmath}
\usepackage{amssymb}

\usepackage{color}
\usepackage{tikz}

\journal{Wave Motion}

\newcommand{\rkl}[1]{\left( #1 \right)}                
\newcommand{\ekl}[1]{\left[ #1 \right]}

\newcommand{\visc}{\eta}
\newcommand{\volvisc}{\eta'}

\newcommand{\markup}[1]{#1}

\begin{document}

\begin{frontmatter}



\title{A weakly nonlinear wave equation for damped acoustic waves with thermodynamic non-equilibrium effects}


\author[MS]{M. Scholle}

\address[MS]{Heilbronn University, Institute for Flow in Additively Manufactured Porous Media, D-74081 Heilbronn, Germany; markus.scholle@hs-heilbronn.de}
%
\begin{abstract}
The problem of propagating nonlinear acoustic waves is considered; the solution to which, both with and without damping, having been obtained to-date starting from the Navier-Stokes-Duhem equations together with the continuity and thermal conduction equation. The novel approach reported here adopts instead, a \emph{discontinuous Lagrangian approach}, i.e.\ from Hamilton's principle together with a discontinuous Lagrangian for the case of a general viscous flow. 
It is shown that ensemble averaging of the equation of motion resulting from the Euler-Lagrange equations, under the assumption of irrotational flow, leads to a weakly nonlinear wave equation for the velocity potential: in effect a generalisation of Kuznetsov's well known equation with an additional term due to thermodynamic non-equilibrium effects.
\end{abstract}


\begin{highlights}
\item Viscous flow with thermal conduction is deducible from a discontinuous Lagrangian
\item Non-classical effects occur beyond thermodynamic equilibrium
\item By ensemble averaging a non-classical equation of motion is derived
\item In the irrotational and weakly nonlinear case a generalised Kuznetsov equation is obtained
\end{highlights}


\end{frontmatter}


\section{Introduction}

The research field of nonlinear acoustics resides between those of fluid mechanics, related to the macroscopic movement of fluid matter, and linear acoustics involving the propagation of small disturbances without significant material movement.  Consequently, classical wave theory has emerged from the underlying continuum theories of fluid mechanics and thermodynamics together with the two basic assumptions that the fluid flow is both compressible and irrotational.
In the classical theory of acoustic waves the relevant equations of motion are also consistently linearised with both viscosity and heat conduction neglected; an adequate approximation for small amplitudes and propagation over short distances. However, in the case of large amplitudes or larger distances damping and nonlinear effects become relevant requiring, in effect, a solution to the full set of fundamental equations of fluid dynamics -- the Navier-Stokes-Duhem equations -- together with the continuity and thermal conduction equations, \cite{olsson2013transport,belevich2017classical,Landau6}, which is very mathematically challenging. As an acceptable alternative \emph{weakly nonlinear models} have been established by considering only the dominant nonlinear contributions, leading typically to a reduced description in terms of a single PDE. In this respect, \cite{Lamb1910the}'s early work has been followed by numerous contributions from other authors such as \cite{lighthill1956viscosity,doi:10.1121/1.410434,ockendon_ockendon_falle_2001,10.1093/qjmam/hbm017,10.1093/qjmam/hbu023} and the many others listed in the chronology of \citet{JORDAN2016127}'s review article.

Despite their variety, weakly nonlinear wave theories are commonly developed from the same classical fluid equations of motion under the assumption of a local thermodynamic equilibrium. Alternatively, it is possible to start from a variational principle for viscous flow since, as is well known, the use of Hamilton's principle is ideally suited to, for example, the field of conservative Newtonian mechanics. Contrary to this, in continuum theories in general and in the viscous flow theory in particular the formulation of variation principles does not follow a classical construction scheme, i.e.\ the Lagrangian cannot simply be considered as the difference between kinetic and potential energy, as analysed in detail by \cite{dualidad}. Accordingly, 
after \cite{clebsch}'s early success with regard to inviscid barotropic flows, a century passed before \cite{Lin} and later \cite{SelWhith} found a generalisation that included thermal degrees of freedom for reversible processes.
 
By reinterpreting the thermal degrees of freedom occurring in \cite{SelWhith}'s Lagrangian in terms of the complex field of thermal excitation, originally introduced by \cite{ISI:A1989EN89700005,anthony1990phenomenological,Anthony}, \citet{Scholle160447} suggested a Lagrangian for incompressible flow with an additional \emph{discontinuous} term taking viscosity into account and containing an additional parameter $\omega_0$.
By careful analysis it is proven that the dynamics resulting from Hamilton's principle can consistently be interpreted as a generalisation of the theory of viscous flow towards thermodynamic non-equilibrium, with the parameter $\omega_0$ being the relaxation rate, giving rise to recovery of the well-known Navier-Stokes equations and the balance of inner energy when applying the limit $\omega_0\to\infty$ to the resulting equations of motion. 

The above \emph{discontinuous Lagrangian approach} has been progressively generalised and better understood: beginning with the inclusion of compressibility, \cite{doi:10.1098/rsos.181595} exposed conceptual parallels to the stochastic calculus of variations, see \cite{w12030864}. By linearisation of the equations of motion, they derived a discontinuous PDE for damped planar acoustic waves. By ensemble averaging \cite{SCHOLLE2020102636} obtained the continuous wave equation:  
\begin{equation}
 \square\varepsilon+\frac{\bar\nu}{a_0^{2}}\left\{1+\frac{\pi^{2}}{2\omega_{0}^{2}}\partial_{t}^{2}\right\}\partial_{t} \partial_x^{2}\varepsilon = 0\,. \label{PDE_epsilon1D}
\end{equation}
for the \markup{condensation} $\varepsilon$, with $ \square$ the d'Alembertian operator, $a_0$ the speed of sound, $\bar\nu$ the diffusivity of sound with the aforementioned additional parameter $\omega_0$ giving rise to the additional term $\frac{\pi^{2}}{2\omega_{0}^{2}}\partial_{t}^{3} \partial_x^{2}\varepsilon$, when compared to the classical linear theory, due to thermodynamic non-equilibrium.

This paper constucts a generalisation of Eq.\ \eqref{PDE_epsilon1D} towards a corresponding weakly nonlinear wave equation. For convenience, viscous damping only is taken into account. Firstly,  the discontinuous Lagrangian approach is revisited in Sec.~\ref{sec:general}, but in contrast to the prior work of \cite{SCHOLLE2020102636} a more detailed analysis is carried out, leading to a classification of non-classical terms into two different categories -- on the one hand to contributions resulting directly from deviations from thermodynamic equilibrium, and on the other hand indirect contributions due to small fluctuations around thermodynamic equilibrium. Only the latter are considered for the theory of nonlinear acoustic waves developed in Sec.~\ref{sec:acoustics}, assuming irrotational flow. This initially leads to a fully non-linear theory comprised of two PDEs, which can subsequently be reduced to a single PDE by neglecting cubic terms. Conclusions are drawn in Sec.~\ref{sec:outlook} together with correspondingly potential further research avenues.

\section{The discontinuous Lagrangian approach}
\label{sec:general}

In the classic theory of fluid flow balance equations are considered, comprising the continuity equation (mass balance), the Navier-Stokes equations momentum balance) and an appropriate thermal conduction-convection equation (partial energy balance), see e.g.\ \cite{belevich2017classical,Landau6}. In contrast, an alternative approach is elaborated by \cite{doi:10.1098/rsos.181595}, starting from a variational principle with a discontinuous Lagrangian for the respective continuum and applying ensemble averaging to the resulting equations of motion.

\subsection{A discontinuous Lagrangian for compressible viscous flow}
\label{sec:theLagrangian}

For the particular case of compressible viscous flow without thermal conductivity, the following Lagrangian was proposed by \citet{doi:10.1098/rsos.181595}:
\begin{align}
 \ell =& -\varrho\left[\mathrm{D}_t\Phi +\alpha\mathrm{D}_t\beta +\frac{1}{\omega_0}\Im\left(\bar\chi\mathrm{D}_t\chi\right) -\frac{{\vec{u}}^{\,2}}{2} + e\left(\varrho,s\right)
\right]\nonumber\\
 &+\frac{1}{\mathrm{i}\omega_0}\ln\sqrt{\frac{\bar\chi}{\chi}}  \left[\eta\mathrm{tr}{\underline{D}}^{2}+\frac{\eta'}{2}\rkl{\nabla\cdot\vec{u}}^{2}\right]\,, \label{l_Scholle2015}
\end{align}
where $e=e\left(\varrho,s\right)$ is the specific inner energy of the fluid depending on the mass density $\varrho$ and the specific entropy, the latter \markup{given by the non-classical\footnote{Different from the classical expression for the specific entropy, e.g.\ in perfect gases, see \cite{JORDAN2016127}, Eq. (19), the thermal excitation $\chi$ takes on the role of the decicive thermodynamic state variable.} expression:
\begin{equation}
 s = c_{p0}\ln\left(\frac{\bar\chi\chi}{c_{p0}T_0}\right) \,, \label{def:s}
\end{equation}
in terms of the complex--valued field thermal excitation $\chi$ originally proposed by \citet{Anthony}, with a reference temperature $T_0$, a reference mass density $\varrho_0$ and the reference specific heat:
\begin{equation}
 c_{p0} := c_p\left(\varrho_0,T_0\right) \,,
\end{equation}
defined as the specific heat for constant pressure at the reference state. $\mathrm{tr}$ denotes the trace of a tensor, $\vec u$ the velocity field, $\text{D}_t=\partial/\partial t+\vec u\cdot\nabla$ the material time derivative, and:
\begin{equation}
 \underline{D}=\frac{1}{2}\ekl{\nabla\otimes\vec{u}+\rkl{\nabla\otimes\vec{u}}^{t}}, \label{shear-rate}
\end{equation}
is the shear rate tensor. The superscript $t$ indicates the transpose of a tensor} and the symbol $\otimes$ the dyadic product. The two coefficients, the shear viscosity, $\visc$, and the dilatational viscosity (Lam\'e's first parameter), $\volvisc$, are assumed to be constant. The list of fundamental fields is completed by the Clebsch variables $\Phi$, $\alpha$ and $\beta$.

The Lagrangian \eqref{l_Scholle2015} consists of two different parts: the first line is essentially the well-known Lagrangian of \citet{Lin,SelWhith} for reversible and therefore inviscid flow on decomposing the thermal excitation into its modulus and its argument, while the terms in the second line take both shear viscosity and volume viscosity into account. A striking feature is its discontinuity due to the logarithmic term, $\ln\sqrt{{\bar\chi}/{\chi}}$. The additional parameter $\omega_0$ occurring in \eqref{l_Scholle2015} has, according to \cite{Scholle160447}, the meaning of a thermodynamic relaxation rate and therefore no equivalent in the classical theory. \markup{Its reciprocal multiplied by $2\pi$, i.e.\ $2\pi/\omega_0$ can according to \cite{doi:10.1098/rsos.181595,SCHOLLE2020102636} be interpreted as a thermodynamic relaxation time, in relation to phenomena away from thermodynamic equilibrium and beyond the continuum hypothesis, in particular Brownian molecular motion, causing according to \citet{PhysRevLett.108.160601} a ``momentum transfer deficit''. In \cite{Scholle160447} the role of Brownian motion for viscosity is emphasised.}


It is important to point out that there is a long-term, on-going debate about the global existence of Clebsch variables; accordingly the interested reader is referred to the recent review paper by \citet{w12051241} for details. Since for the subsequent application to nonlinear acoustics, only irrotational flows are considered, there is no need here to examine this problem further.

\subsection{Euler-Lagrange equations}
\label{app:EL}

The associated Euler-Lagrange equations are obtained by variation of the action integral:
\begin{equation}
 \delta \int\limits_{t_1}^{t_2}\iiint\limits_V \ell\left(\psi_i,{\dot\psi}_i,\nabla\psi_i\right) \text{d}V\text{d}t =0\,, \label{Hamiltonsprinciple}
\end{equation}
w.r.t.\ the fields $\left(\psi_i\right)=\left(\Phi,\alpha,\beta,\varrho,\vec{u},\chi,\bar{\chi}\right)$ and their first order spatial and temporal derivatives, $\nabla\psi_i$ and $\dot\psi_i=\partial_t\psi_i$, for fixed values at initial and final time, $t_{1,2}$.
For the Lagrangian \eqref{l_Scholle2015}, they result, after minor manipulations, in: 
\begin{align}
\partial_{t}\varrho+\nabla\cdot\left(\varrho\vec{u}\right) & =0\,.\label{delta_Phi_compr}\\
\mathrm{D}_{t}\beta & =0\,,\label{delta_alpha_compr}\\
\mathrm{D}_{t}\alpha & =0\,,\label{delta_beta_compr}\\
\mathrm{D}_{t}\Phi+\alpha\mathrm{D}_{t}\beta+\frac{1}{\omega_{0}}\Im\left(\bar{\chi}\mathrm{D}_{t}\chi\right) & =\frac{{\vec{u}}^{\,2}}{2}-h\,,\label{delta_rho_compr}\\
\vec{u} & =\frac{\vec{p}}{\varrho}-\frac{1}{\varrho\omega_{0}}\nabla\cdot\left(\mathrm{i}\ln\sqrt{\frac{\bar{\chi}}{\chi}}\underline{R}\right)\,,\label{delta_u_compr}\\
\text{D}_{t}\chi & =\frac{P_{\text{diss}}}{4\varrho\bar{\chi}}-\frac{\text{i}\omega_{0}}{\bar{\chi}}c_{p0}T\,,\label{delta_barchi_compr}
\end{align}
utilising the thermodynamic relation $T= {\partial e}/{\partial s}$, with the friction tensor $\underline{R}$, the (canonical) momentum density $\vec{p}$ (resulting via Noether's theorem according to invariance) of the Lagrangian w.r.t.\ spatial translations, the specific enthalpy $h$ and the dissipation rate $P_{\text{diss}}$ given by:
\begin{align*}
\underline{R} & :=2\eta\underline{D}+\eta'\nabla\cdot\vec{u}\,\underline{1} \,, \\
\vec{p} & :=\varrho\left[\nabla\Phi+\alpha\nabla\beta+\frac{1}{\omega_{0}}\Im\left(\bar{\chi}\nabla\chi\right)\right]\,, \\
h & :=e+\varrho\frac{\partial e}{\partial\varrho}\,, \\
P_{\text{diss}} & :=2\eta\mathrm{tr}{\underline{D}}^{2}+\eta'\left(\nabla\cdot\vec{u}\right)^{2},
\end{align*}
respectively.

Equation \eqref{delta_Phi_compr} is that of continuity, (\ref{delta_alpha_compr}--\ref{delta_rho_compr}) are evolution equations for the Clebsch variables, while the evolution of the thermal excitation is given by \eqref{delta_barchi_compr}. As a non-classical feature of the present approach, the disparity between momentum density $\vec p$ and mass flux density $\varrho\vec u$ is revealed by \eqref{delta_u_compr}, which according to \cite{Scholle160447} can also be understood as a manifestation of a thermodynamic non-equilibrium.

\subsection{Discontinuous equation of motion}

Subsequently, the corresponding equation of motion is derived utilising the identity:
\begin{align*}
\left\{ \mathrm{D}_{t}+\nabla\otimes\vec{u}\right\} \frac{\vec{p}}{\varrho}&=\nabla\left[\mathrm{D}_{t}\Phi+\alpha\mathrm{D}_{t}\beta+\frac{1}{\omega_{0}}\Im\left(\bar{\chi}\mathrm{D}_{t}\chi\right)\right]\\
 & +\mathrm{D}_{t}\alpha\nabla\beta-\mathrm{D}_{t}\alpha\nabla\beta-\frac{2}{\omega_{0}}\Im\left(\mathrm{D}_{t}\chi\nabla\bar{\chi}\right) ,
\end{align*}
for the specific momentum. By inserting the Euler-Lagrange equations (\ref{delta_alpha_compr}--\ref{delta_rho_compr}) and \eqref{delta_barchi_compr}, one obtains:
\begin{align*}
\left\{ \mathrm{D}_{t}+\nabla\otimes\vec{u}\right\} \frac{\vec{p}}{\varrho} &  =\nabla\left[\frac{{\vec{u}}^{\,2}}{2}-h\right]-\frac{1}{\omega_{0}}\Im\left(\left[\frac{P_{\text{diss}}}{2\varrho}-2\text{i}\omega_{0}c_{p0}T\right]\frac{\nabla\bar{\chi}}{\bar{\chi}}\right) ,\\
 & =\nabla\left[\frac{{\vec{u}}^{\,2}}{2}-h\right]-\frac{P_{\text{diss}}}{2\varrho\omega_{0}}\Im\left(\frac{\nabla\bar{\chi}}{\bar{\chi}}\right)+\underbrace{2c_{p0}T\Re\left(\frac{\nabla\bar{\chi}}{\bar{\chi}}\right)}_{T\nabla s} ,\\
 &=\left(\nabla\otimes\vec{u}\right)\vec{u}\underbrace{-\nabla h+T\nabla s}_{-\nabla p/\varrho}-\frac{P_{\text{diss}}}{\varrho\omega_{0}}\Im\left(\frac{\nabla\bar{\chi}}{\bar{\chi}}\right) .
\end{align*}
Finally, the equation of motion is obtained by applying the material time
derivative $\mathrm{D}_{t}$ to \eqref{delta_u_compr}:
\begin{align*}
\mathrm{D}_{t}\vec{u} & =\mathrm{D}_{t}\left(\frac{\vec{p}}{\varrho}\right)-\frac{1}{\omega_{0}}\mathrm{D}_{t}\left[\frac{1}{\varrho}\nabla\cdot\left(\mathrm{i}\ln\sqrt{\frac{\bar{\chi}}{\chi}}\underline{R}\right)\right] ,\\
 & =\left(\nabla\otimes\vec{u}\right)\underbrace{\left[\vec{u}-\frac{\vec{p}}{\varrho}\right]}_{-\frac{\nabla\cdot\left(\mathrm{i}\ln\sqrt{\frac{\bar{\chi}}{\chi}}\underline{R}\right)}{\omega_{0}\varrho}}-\frac{\nabla p}{\varrho}-\frac{P_{\text{diss}}}{2\varrho\omega_{0}}\Im\left(\frac{\nabla\bar{\chi}}{\bar{\chi}}\right)-\mathrm{D}_{t}\left[\frac{1}{\omega_{0}\varrho}\nabla\cdot\left(\mathrm{i}\ln\sqrt{\frac{\bar{\chi}}{\chi}}\underline{R}\right)\right]\,,
\end{align*}
resulting finally in:
\begin{equation}
\mathrm{D}_{t}\vec{u}=-\frac{\nabla p}{\varrho}+\frac{P_{\text{diss}}}{2\varrho\omega_{0}}\Im\left(\frac{\nabla\chi}{\chi}\right)-\left\{ \mathrm{D}_{t}+\nabla\otimes\vec{u}\right\} \left[\frac{1}{\omega_{0}\varrho}\nabla\cdot\left(\mathrm{i}\ln\sqrt{\frac{\bar{\chi}}{\chi}}\underline{R}\right)\right]\,. \label{eom}
\end{equation}
\citet{Scholle160447} showed that this equation reproduces the Navier-Stokes equations supplemented by fluctuating additional terms as thermodynamic non-equilibrium contributions.

\subsection{Time scale separation}

We consider as a reference state a fluid at rest, $\vec u=0$, with constant temperature $T=T_{0}$ and entropy $s=s_{0}=0$.
Under these conditions the evolution equation \eqref{delta_barchi_compr} for the thermal excitation has the solution:
\begin{equation}
 \chi=\chi_{0}(t):=\sqrt{c_{p0}T_{0}}\exp\left(-\text{i}\left[\varphi_{0}-\omega_{0}t\right]\right)\,,
\end{equation}
with phase shift $\varphi_0=\text{const}$.

Next we consider dynamic processes deviating only slightly from the above reference state and make the following substitution:
\begin{equation}
\chi=\chi_{0}\exp\left(\zeta\right)\,, \label{def:zeta}
\end{equation}
that factorises the thermal excitation into the product of an equilibrium quantity $\chi_{0}$ varying rapidly with time 
and the slowly varying relative thermal excitation $\zeta$; the meaning of the latter
becomes apparent by the definition \eqref{def:s} of the entropy:
\begin{equation}
 s = c_{p0}\ln\left(\frac{\bar\chi\chi}{c_{p0}T_0}\right)=c_{p0}\left[\ln\exp\left(\bar{\zeta}\right)+\ln\exp\left(\zeta\right)\right]=2c_{p0}\Re\zeta \,, \label{new:s}
\end{equation}
and by the relation: 
\begin{align}
 \mathrm{i}\ln\sqrt{\frac{\bar{\chi}}{\chi}}  &=\mathrm{i}\ln\left(\exp\left(-\text{i}\left[\varphi_{0}-\omega_{0}t\right]\right)\right)+\frac{\mathrm{i}}{2}\left[\ln\exp\left(\bar{\zeta}\right)-\ln\exp\left(\zeta\right)\right] ,\nonumber\\
  &=S\left(\varphi_{0}-\omega_{0}t\right)+\Im\zeta \,, \label{phi-split}
\end{align}
with the sawtooth function $S(x)=\mathrm{i}\ln\left(\exp\left(-\text{i}x\right)\right)=x-2\pi\left\lfloor \frac{x+\pi}{2\pi}\right\rfloor$,
splitting the thermal phase $\mathrm{i}\ln\sqrt{{\bar{\chi}}/{\chi}}$ into a rapidly varying part $S\left(\varphi_{0}-\omega_{0}t\right)$  physically associated to fluctuations according to \cite{Scholle160447} and a slowly varying part $\Im\zeta$. The latter vanishes for a perfect thermodynamic equilibrium and is assumed to be small, $\left|\zeta\right|\ll1$, for processes not far away from thermodynamic equilibrium.

By introducing \eqref{def:zeta} to the equation of motion \eqref{eom}, the viscous terms on the right hand side decompose according to:
\begin{equation}
 \mathrm{D}_{t}\vec{u}=-\frac{\nabla p}{\varrho}
  +\frac{\vec{f}_{1}}{2\omega_{0}}
  -\left\{ \mathrm{D}_{t}+\nabla\otimes\vec{u}\right\} \left[S\left(\varphi_{0}-\omega_{0}t\right)\frac{\nabla\cdot\underline{R}}{\omega_{0}\varrho}\right] ,\label{eom_split}
\end{equation}
where 
\begin{equation}
 \vec{f}_{1}  :=\frac{P_{\text{diss}}}{\varrho}\nabla\Im\zeta-\left\{ \mathrm{D}_{t}+\nabla\otimes\vec{u}\right\} \left[\frac{2}{\varrho}\nabla\cdot\left(\Im\zeta\underline{R}\right)\right] ,
\end{equation}
represents all terms depending on the deviation from thermodynamic equilibrium, $\Im\zeta$, while the last term in \eqref{eom_split} is rapidly fluctuating due the occurrence of $S\left(\varphi_{0}-\omega_{0}t\right)$.

\subsection{Averaged equation of motion}

The rapidly fluctuating part $S\left(\varphi_{0}-\omega_{0}t\right)$ of the thermal phase does not become manifest for an external observer, who can only measure average values. We therefore apply an averaging to the equation of motion, with reference to the ensemble averaging used by \cite{SCHOLLE2020102636}, which is essentially averaging over the phase shift from $-\pi$ to $\pi$:
\[
\left<\cdots\right>:=\frac{\omega_{0}^{2}}{2\pi}\int\limits _{-\pi}^{+\pi}\left[\cdots\right]\text{d}\varphi_{0}\,.
\]
For convenience we compute the equivalent material time average: let $f=f\left(\vec x,t\right)$ be an arbitrary field. Then, applying a Taylor expansion, the average of $S\left(\varphi_{0}-\omega_{0}t\right)f$ is computed as:
\begin{align*}
\left<S\left(\varphi_{0}-\omega_{0}t\right)f\right> & =\frac{\omega_{0}}{2\pi}\int\limits _{-\pi/\omega_{0}}^{+\pi/\omega_{0}} (-\omega_{0}\tau) f\left(\vec{x}+\vec{u}\tau,t+\tau\right)\text{d}\tau ,\\
 & =-\frac{\omega_{0}^{2}}{2\pi}\int\limits _{-\pi/\omega_{0}}^{+\pi/\omega_{0}}\tau\left[f\left(\vec{x},t\right)+\text{D}_{t}f\tau+\text{D}_{t}^{2}f\frac{\tau^{2}}{2}+{\cal O}\left(\tau^{3}\right)\right]\text{d}\tau ,\\
 & =-\frac{\omega_{0}^{2}}{2\pi}\left[0+\text{D}_{t}f\frac{2\pi^{3}}{3\omega_{0}^{3}}+0+{\cal O}\left(\omega_{0}^{-5}\right)\right]
  =-\frac{\pi^{2}}{3\omega_{0}}\text{D}_{t}f+{\cal O}\left(\omega_{0}^{-3}\right)\,,
\end{align*}
while the average of the material time derivative of the same expression gives:
\begin{align*}
\left\langle \text{D}_{t}\left[S\left(\varphi_{0}-\omega_{0}t\right)f\right]\right\rangle  & =\frac{\omega_{0}}{2\pi}\int\limits _{-\pi/\omega_{0}}^{+\pi/\omega_{0}}\frac{\text{d}}{\text{d}\tau}\left[\left(-\omega_{0}\tau\right)f\left(\vec{x}+\vec{u}\tau,t+\tau\right)\right]\omega_{0}\text{d}\tau ,\\
 & =-\frac{\omega_{0}^{2}}{2\pi}\left.\left[\tau f\left(\vec{x}+\vec{u}\tau,t+\tau\right)\right]\right|_{-\pi/\omega_{0}}^{+\pi/\omega_{0}} ,\\
 & =-\frac{\omega_{0}}{2}\left[f\left(\vec{x}+\vec{u}\textstyle\frac{\pi}{\omega_0},t+\textstyle\frac{\pi}{\omega_{0}}\right) +f\left(\vec{x}-\vec{u}\textstyle\frac{\pi}{\omega_0},t-\textstyle\frac{\pi}{\omega_{0}}\right)+{\cal O}\left(\omega_{0}^{-5}\right)\right] ,\\
 & =-\omega_{0}f\left(\vec{x},t\right)-\frac{\pi^{2}}{2\omega_{0}}\text{D}_{t}^{2}f\left(\vec{x},t\right)+{\cal O}\left(\omega_{0}^{-3}\right).
\end{align*}
The above averaging applied to the fluctuating term in the equation of motion \eqref{eom_split} leads to:
\begin{align}
&-\left\langle \left\{ \mathrm{D}_{t}+\nabla\otimes\vec{u}\right\} \left[S\left(\varphi_{0}-\omega_{0}t\right)\frac{\nabla\cdot\underline{R}}{\omega_{0}\varrho}\right]\right\rangle \label{<friction>} \\
 & =\frac{1}{\varrho}\nabla\cdot\underline{R}+\frac{\pi^{2}}{2\omega_{0}^{2}}\text{D}_{t}^{2}\left(\frac{1}{\varrho}\nabla\cdot\underline{R}\right)+\frac{\pi^{2}}{3\omega_{0}^{2}}\text{D}_{t}\left(\frac{1}{\varrho}\nabla\otimes\vec{u}\,\nabla\cdot\underline{R}\right)+{\cal O}\left({\omega_{0}^{-3}}\right)\,.\nonumber
\end{align}
With $\nabla\cdot\underline{R}=\eta\nabla^2\vec u+\left(\eta+\eta'\right)\nabla\left(\nabla\cdot\vec u\right)$, the averaged equation of motion is finally:
\begin{equation}
  \mathrm{D}_{t}\vec{u}=-\frac{\nabla p}{\varrho} +\frac{\eta}{\varrho}\nabla^2\vec u+\frac{\eta+\eta'}{\varrho}\nabla\left(\nabla\cdot\vec u\right) 
    +\frac{\vec{f}_{1}}{2\omega_{0}} +\frac{\pi^{2}\vec{f}_{2}}{2\omega_{0}^{2}}+{\cal O}\left(\omega_{0}^{-3}\right) , \label{<eom>}
\end{equation}
with the abbreviation:
\begin{equation}
\vec{f}_{2}  :=\text{D}_{t}\left\{ \text{D}_{t}+\frac{2}{3}\nabla\otimes\vec{u}\right\} \left[\frac{\eta}{\varrho}\nabla^2\vec u+\frac{\eta+\eta'}{\varrho}\nabla\left(\nabla\cdot\vec u\right)\right]\,,
\end{equation}
resulting in the Navier-Stokes-Duhem equation with additional non-classical contributions. Although this result has already been discussed in a broader sense by \citet{doi:10.1098/rsos.181595}, the more detailed analysis carried out above leads to a classification of non-classical terms into two different categories, namely on the one hand the contributions summarised under $\vec f_1$, which result directly from deviation from thermodynamic equilibrium, and on the other hand the contributions summarised under $\vec f_2$, which result indirectly as mean value of forces due to fluctuations around equilibrium.

\section{Application to nonlinear acoustics}
\label{sec:acoustics}

In what follows, the averaged equation of motion \eqref{<eom>} is used instead of the classical Navier-Stokes-Duhem equation as the starting point for the derivation of a weakly non-linear evolution equation for acoustic waves.

\subsection{Basic assumptions}

Four essential assumptions are required. The first concerns the state equation for the pressure: 
\begin{equation}
 \frac{\partial p}{\partial\varrho} = a_0^2\left[1+\frac{B}{A}\frac{\varrho-\varrho_0}{\varrho_0}\right] ,  \label{assump4p}
\end{equation}
with small-signal sound speed $a_0$ from linear theory and containing the nonlinearity parameter $B/A$ of the respective fluid, \cite{parameterBoverA,rasmussen2008analytical}. \markup{Typical values of the nonlinearity parameter are according to \cite{parameterBoverA} for instance $0.4$ for air at 20\textdegree{} and $5.0$ for water at 20\textdegree{}. For linear alkanes (from pentane to octane) typical values are according to \cite{article} between 6.5 and 10.}
Secondly, the diffusivity of sound\markup{, which for a non-thermally conducting fluid reads:}
\begin{equation}
 \bar{\nu}:=\frac{2\eta+\eta'}{\markup{\varrho_0}}\, .
\end{equation}
Furthermore, sound waves are regarded as irrotational, $\nabla\times\vec u=\vec 0$, implying:
\begin{equation}
 \vec u = \nabla\Phi\,. \label{assump4u}
\end{equation}
We finally consider in the averaged equation of motion \eqref{<eom>} only the non-classical contributions due to fluctuations $\vec f_2$ while assuming $\Im\zeta=0$, implying that deviations from thermodynamic equilibrium do not become manifest directly. As a consequence:
\begin{equation}
 \vec{f}_{1}=\vec{0}\,. \label{assump4zeta}
\end{equation}

\subsection{Fully nonlinear equations}

Based on the assumptions highlighted above, the material acceleration and the specific friction force become gradient fields:
\begin{align}
  \mathrm{D}_{t}\vec{u} &=\nabla\left[\partial_{t}\Phi+\frac{1}{2}\left(\nabla\Phi\right)^{2}\right] ,\\
  \frac{\eta}{\varrho}\nabla^{2}\vec{u}+\frac{\eta+\eta'}{\varrho}\nabla\left(\nabla\cdot\vec{u}\right) & =\bar{\nu}\nabla\left(\nabla^{2}\Phi\right).
\end{align}
For \markup{convenience the condensation $\varepsilon:=\ln\left(\varrho/\varrho_0\right)$ is introduced\footnote{In many papers $(\varrho-\varrho_0)/\varrho_0$ is defined as condensation, which results from \eqref{def:condensation} via Taylor expansion.}, allowing in turn to express the density as}:
\begin{equation}
 \varrho=\varrho_{0}\exp\left(\varepsilon\right) \label{def:condensation}
\end{equation}
\markup{
and to rewrite the specific pressure gradient} as a gradient field:
\begin{align}
\frac{\nabla p}{\varrho} & =\frac{\partial p}{\partial\varrho}\frac{\nabla\varrho}{\varrho}
=a_0^{2}\left[1+\frac{B}{A}\left(\exp\left(\varepsilon\right)-1\right)\right]\nabla\varepsilon \nonumber\\ 
&=a_0^{2}\nabla\left[\varepsilon+\frac{B}{A}\left(\exp\left(\varepsilon\right)-1-\varepsilon\right)\right].
\end{align}
In contrast, the non-classical contributions in the averaged equation of motion:
\begin{equation}
 \vec{f}_{2} =\bar\nu\left\{ \partial_{t}+\nabla\Phi\cdot\nabla\right\} \left\{ \partial_{t}+\nabla\Phi\cdot\nabla+\frac{2}{3}\nabla\otimes\nabla\Phi\right\} \nabla\left(\nabla^{2}\Phi\right)
\end{equation}
do not result exactly in a gradient field. However, since they contribute to the equation of motion only with factor $\bar\nu \omega_0^{-2}$, it is legitimate to consider only the leading term by linear approximation:
\begin{equation}
 \vec{f}_{2} \approx \bar\nu\partial_{t}^{2}\nabla \left(\nabla^{2}\Phi\right)
\end{equation}
delivering the desired gradient field. 

Based on the above, the averaged equation of motion \eqref{<eom>} takes the integrable form:
\[
\nabla\left[\partial_{t}\Phi+\frac{1}{2}\left(\nabla\Phi\right)^{2}+a_0^{2}\left[\varepsilon+\frac{B}{A}\left(\exp\left(\varepsilon\right)-1-\varepsilon\right)\right]-\bar{\nu}\left\{1+\frac{\pi^{2}}{2\omega_{0}^{2}}\partial_{t}^{2}\right\}\nabla^{2}\Phi\right]=0\,,
\]
which after integration leads to:
\[
\partial_{t}\Phi+\frac{1}{2}\left(\nabla\Phi\right)^{2}+a_0^{2}\left[\varepsilon+\frac{B}{A}\left(\exp\left(\varepsilon\right)-1-\varepsilon\right)\right]-\bar{\nu}\left\{1+\frac{\pi^{2}}{2\omega_{0}^{2}}\partial_{t}^{2}\right\}\nabla^{2}\Phi=F\left(t\right)\,.
\]
By gauging of the potential, $\Phi\to\Phi+\int F(t)\text{d}t$, the integration function $F(t)$ can be absorbed, giving:
\begin{equation}
\partial_{t}\Phi+\frac{1}{2}\left(\nabla\Phi\right)^{2}+a_0^{2}{\left[\varepsilon+\frac{B}{A}\left(\exp\left(\varepsilon\right)-1-\varepsilon\right)\right]}-\bar{\nu}\left\{ 1+\frac{\pi^{2}}{2\omega_{0}^{2}}\partial_{t}^{2}\right\} \nabla^{2}\Phi=0\, .\label{eq:eom_ac}
\end{equation}
While, on the other hand, the set of equations is completed by the continuity equation:
\begin{equation}
0=\frac{\mathrm{D}_{t}\varrho}{\varrho}+\nabla\cdot\vec{u}=\mathrm{D}_{t}\varepsilon+\nabla^{2}\Phi=\left\{ \partial_{t}+\nabla\Phi\cdot\nabla\right\} \varepsilon+\nabla^{2}\Phi\,,\label{eq:cont_ac}
\end{equation}
resulting in a set of two PDEs, \eqref{eq:eom_ac} and \eqref{eq:cont_ac}, for the two unknown fields $\Phi$ and $\varepsilon$.

\subsection{Reduction to one weakly nonlinear equation}

The two equations \eqref{eq:eom_ac} and \eqref{eq:cont_ac} reduce to only one equation by Taylor expansion and omitting terms of cubic and higher order \markup{with respect to the condensation and the derivatives of the potential. A second and independent assumption is weak damping of the waves, justifying to consider damping terms only in leading order}.

First, by considering $\exp\left(\varepsilon\right)-1-\varepsilon\approx \varepsilon^2/2+{\cal O}\left(\varepsilon^{3}\right)$, Eq.\ \eqref{eq:eom_ac} becomes a simple quadratic equation w.r.t\ $\varepsilon$:
\begin{align}
& \varepsilon^{2}+\frac{2A}{B}\varepsilon+\frac{2A}{Ba_0^{2}}q=0\,,\label{quadratic-eq}\\
& q :=  \partial_{t}\Phi+\frac{1}{2}\left(\nabla\Phi\right)^{2}-\bar{\nu}\left\{ 1+\frac{\pi^{2}}{2\omega_{0}^{2}}\partial_{t}^{2}\right\} \nabla^{2}\Phi\,,
\end{align}
\markup{where the the abbreviation $q$ is used for convenience.} 

\markup{Second, a small velocity\footnote{This is usually assumed for acoustic problems.} $\vec u=\nabla\Phi$ is assumed, implying via the continuity equation \eqref{eq:cont_ac} and the non-classical Bernoulli equation \eqref{eq:eom_ac} that the spatial and temporal derivatives of the velocity potential $\Phi$ and consequently also the expression $q$ are small and of the same order as the condensation $\varepsilon$.
Therefore, the two solutions of \eqref{quadratic-eq}} read, on applying a Taylor expansion:
\begin{equation}
 \varepsilon  =-\frac{A}{B}\pm\frac{A}{B}\sqrt{1-\frac{2Bq}{Aa_0^{2}}} \approx -\frac{A}{B}\pm\frac{A}{B}\left[1-\frac{Bq}{Aa_0^{2}}-\frac{B^2q^2}{2A^2a_0^{4}}\right] +{\cal O}\left(q^{3}\right).
\end{equation}
Since in the reference state, $\Phi=0$, the \markup{condensation} has to vanish, the upper sign gives the right solution.

\markup{Third, we consider the square
\begin{align*}
 q^2 &= \left(\partial_{t}\Phi-\bar{\nu}\left\{ 1+\frac{\pi^{2}}{2\omega_{0}^{2}}\partial_{t}^{2}\right\} \nabla^{2}\Phi\right)^2\\
         &    +\left(\partial_{t}\Phi-\bar{\nu}\left\{ 1+\frac{\pi^{2}}{2\omega_{0}^{2}}\partial_{t}^{2}\right\} \nabla^{2}\Phi\right)\left(\nabla\Phi\right)^{2}
             +\frac{1}{4}\left(\nabla\Phi\right)^{4}
\end{align*}
and neglect the entire second line of above formula, since it contains only terms of cubic and higher order. Additionally, we consider terms related to damping only in leading order, i.e.\ only linear terms. This also concerns $\bar{\nu} \nabla^{2}\Phi \partial_{t}\Phi$ and $\bar{\nu}^2 \left(\nabla^{2}\Phi \partial_{t}\Phi\right)^2$, leading to the approximation $q^2\approx \left(\partial_{t}\Phi\right)^2$ and therefore to the relation:}
\begin{equation}
 \varepsilon  =-\frac{1}{a_0^{2}}\left[\partial_{t}\Phi+\frac{1}{2}\left(\nabla\Phi\right)^{2}-\bar{\nu}\left\{ 1+\frac{\pi^{2}}{2\omega_{0}^{2}}\partial_{t}^{2}\right\} \nabla^{2}\Phi\right]-\frac{B}{2Aa_0^{4}}\left(\partial_{t}\Phi\right)^{2} \,. \label{sol:epsilon}
\end{equation}
%
Finally, \markup{the desired single equation is obtained on elimination of $\varepsilon$} by inserting \eqref{sol:epsilon} into equation \eqref{eq:cont_ac}. \markup{In the course of this, terms containing derivatives of the velocity potential $\Phi$ of cubic and higher order are again neglected, which leads to the} single PDE:
\begin{equation}
 \square\Phi-\frac{1}{a_0^{2}}\partial_{t}\left[\left(\nabla\Phi\right)^{2}+\frac{B}{2Aa_0^{2}}\left(\partial_{t}\Phi\right)^{2}-\bar\nu\left\{ 1+\frac{\pi^{2}}{2\omega_{0}^{2}}\partial_{t}^{2}\right\} \nabla^{2}\Phi\right] = 0 \label{PDE_nonlinear}
\end{equation}
for the velocity potential $\Phi$, using the following definition of the d'Alembertian: 
\[
\square:=-\frac{1}{a_0^{2}}\partial_{t}^{2}+\nabla^{2}\,.
\]
The nonlinear PDE \eqref{PDE_nonlinear} recovers the well-known \cite{ISI:A1971J209700009} equation in the limit case $\omega_0\to\infty$ and is therefore a generalisation of the latter with a nonclassical supplement $-\bar\nu \omega_{0}^{-2} \pi^{2}\partial_{t}^{2}\nabla^{2}\Phi/2$ to the damping term due to thermodynamic non-equilibrium fluctuations.

\subsection{Linearised equation}

\begin{figure}[htbp]
\begin{center}
\begin{tikzpicture}[>=latex,xscale=7.5,yscale=7.5]
\draw[->] (0,0) -- (.53,0) node[right]{$\frac{\bar\nu}{a_0^2}\omega$}; 
\draw[->] (0,0) -- (0,.43) node[above]{$\frac{\bar\nu}{a_0}\Re k$};
\foreach \x in {0,.1,.2,.3,.4,.5}{\draw[thin] (\x,0.01) -- (\x,0) node[below]{\x};}
\foreach \x in {0,.1,.2,.3,.4}{\draw[thin] (0.01,\x) -- (0,\x) node[left]{{\x}};}
%
\draw[domain=0:1/2] plot (\x,{\x*(1-pi^2*\x^2/18)*cos(atan(\x*(1-pi^2*\x^2/18))/2)/(1+\x^2*(1-pi^2*\x^2/18)^2)^(1/4)});
%
\draw[dashed,domain=0:1/2] plot (\x,{\x*cos(atan(\x)/2)/(1+\x^2)^(1/4)});
%
\color{black}%
\draw[dotted] (0,0) -- (.5,.5);
\end{tikzpicture}
\qquad
\begin{tikzpicture}[>=latex,xscale=7.5,yscale=7.5]
\draw[->] (0,0) -- (.43,0) node[right]{$\frac{\bar\nu}{a_0}\omega$}; 
\draw[->] (0,0) -- (0,.43) node[above]{$\frac{\bar\nu}{a_0}\Im k$};
\foreach \x in {0,.1,.2,.3,.4}{\draw[thin] (\x,0.01) -- (\x,0) node[below]{\x};}
\foreach \x in {0,.02,.04,.06,.08}{\draw[thin] (0.01,\x*5) -- (0,\x*5) node[left]{\x};}
%
\draw[domain=0:1/2] plot (\x,{5*\x*(1-pi^2*\x^2/18)*sin(atan(\x*(1-pi^2*\x^2/18))/2)/(1+\x^2*(1-pi^2*\x^2/18)^2)^(1/4)});
%
\draw[dashed,domain=0:.46] plot (\x,{5*\x*sin(atan(\x)/2)/(1+\x^2)^(1/4)});
\end{tikzpicture}
\caption{Dispersion relation $\Re k\left(\omega\right)$ (left) and attenuation coefficient $\Im k\left(\omega\right)$ (right), plotted versus the angular frequency for the special choice $\omega_0=3a_0^2/\bar\nu$ of the relaxation rate as solid lines. For comparison, the respective relations following from the classical theory ($\omega_0\to\infty$) are shown as dashed lines. The dotted line in the left diagram indicates linear dispersion ($\bar\nu=0$).}
\label{fig:dispersionrelation}
\end{center}
\end{figure}
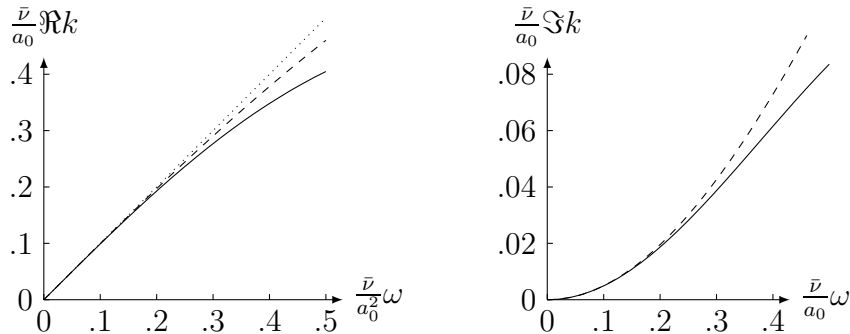

By omitting all nonlinear terms, equation \eqref{PDE_nonlinear} simplifies to:
\begin{equation}
 \square\Phi+\frac{\bar\nu}{a_0^{2}}\left\{ 1+\frac{\pi^{2}}{2\omega_{0}^{2}}\partial_{t}^{2}\right\}\partial_{t} \nabla^{2}\Phi = 0\,. \label{PDE_linear}
\end{equation}
The above equation for damped acoustic waves can alternatively be formulated in terms of the \markup{condensation}, $\varepsilon$, rather than the velocity potential $\Phi$: taking the derivative of the equation w.r.t.\ time and utilising relation \eqref{sol:epsilon}, which in linearised form reads $\partial_{t}\Phi=-a_0^2\varepsilon$, gives:
\begin{equation}
 \square\varepsilon+\frac{\bar\nu}{a_0^{2}}\left\{ 1+\frac{\pi^{2}}{2\omega_{0}^{2}}\partial_{t}^{2}\right\}\partial_{t} \nabla^{2}\varepsilon = 0\, , \label{PDE_epsilon}
\end{equation}
which is of the same analytic form as \eqref{PDE_linear} and a 3D generalisation of the respective one-dimensional wave equation \eqref{PDE_epsilon1D} proposed by \cite{SCHOLLE2020102636}. In the same paper its associated dispersion relation for planar harmonic sound waves with circular frequency $\omega>0$ and wavenumber $k$ has been derived as:
\begin{equation}
  \left[1-\text{i}\frac{\bar\nu\omega}{a_0^2}\left(1-\frac{\pi^2\omega^2}{2\omega_0^2}\right)\right]k^2 =\frac{\omega^2}{a_0^2} \,,\label{disp-relation}
\end{equation}
which differs from the classical one, in particular by the occurrence of the factor $1-{\pi^2\omega^2}/{2\omega_0^2}$.
\markup{The resulting consequences are examined below: considering wave propagation in positive $x$-direction, the solution:
\begin{equation}
 k=\frac{\omega}{a_{0}}\sqrt{\frac{1+\text{i}\frac{\bar{\nu}\omega}{a_{0}^2}\left(1-\frac{\pi^2\omega^2}{2\omega_0^2}\right)}{1+\left(\frac{\bar{\nu}\omega}{a_{0}^2}\right)^{2}\left(1-\frac{\pi^2\omega^2}{2\omega_0^2}\right)^{2}}}\,, \label{k_exact}
\end{equation}
results for the complex wavenumber. It is obvious that the attenuation coefficient, given as the imaginary part of the complex wavenumber, $\Im k$, is reduced for very high angular frequencies $\omega$ compared to the classical theory, due to the factor $1-{\pi^2\omega^2}/{2\omega_0^2}$, as shown in the right diagram of Fig.\ \ref{fig:dispersionrelation}.}
\markup{This can be interpreted physically as follows: if the period of oscillation comes close to the thermodynamic relaxation time, the oscillation is faster than the dissipation process, so that the dissipation is reduced overall. On the other hand the dispersion shown in the left diagram of Fig.\ \ref{fig:dispersionrelation} deviates stronger from linear dispersion as in the classical case.

A striking feature of the complex dispersion relation \eqref{k_exact} for waves propagating in positive $x$-direction is that if the angular frequency $\omega$ exceeds the critical value
\begin{equation}
 \omega_{\text{ c}}:=\frac{\sqrt{2}}{\pi}\omega_{0}\,, \label{omega_c}
\end{equation}
the attenuation coefficient $\Im k$ changes its sign, which physically makes no sense. Thus, in contrast to classical theory, an upper bound $\omega_{\text{ c}}$ for the angular frequency is identified by the dispersion relation \eqref{disp-relation}, above which wave propagation is impossible.

Likewise, by substituting \eqref{omega_c} into \eqref{k_exact}, one obtains a critical wavenumber $k_{\text{c}}=\omega_{\text{ c}}/a_{0}$ and thus a critical wavelength
\begin{equation}
  \lambda_{\text{c}}:=\frac{2\pi}{k_{\text{c}}}=\frac{\sqrt{2}\pi^{2}a_{0}}{\omega_{0}}\,, \label{lambda_c}
\end{equation}
acting as a lower bound for wavelength, i.e.\ $\lambda>\lambda_{\text{c}}$. 
Against the background that the discontinuous Lagrangian approach described in Sect.\ \ref{sec:general} goes beyond the scope of the continuum hypothesis, $\lambda_{\text{c}}$ can logically be assigned to a length scale where deviations from the continuum hypothesis becomes apparent. This suggests estimating $\lambda_{\text{c}}$ by the mean free path length of the molecules of the medium, which in turn allows for an estimation of the lower bound of the relaxation rate $\omega_{0}$ via \eqref{lambda_c}.}

\section{Conclusions and Outlook}
\label{sec:outlook}

Starting from Hamilton's principle with a discontinuous Lagrangian proposed by \citet{doi:10.1098/rsos.181595}, the \markup{general} theory of classical viscous flow is recovered after time scale separation and ensemble averaging, apart from the additional non-classical forces that appear which can be classified into two different categories, namely direct and indirect non-equilibrium contributions, the latter due to small fluctuations around the equilibrium.

Damped nonlinear acoustic waves result from the general theory by assuming irrotational motion, the equation of state \eqref{assump4p} and thermodynamic equilibrium on the larger time scale. The fully nonlinear theory is based on the resulting two PDEs \eqref{eq:eom_ac} and \eqref{eq:cont_ac} for the \markup{condensation} and the velocity potential, respectively. A weakly nonlinear approximation is obtained by omitting all terms of cubic or higher order \markup{and terms related to damping of higher than linear order}, allowing for elimination of the \markup{condensation} and leading to just a single PDE \eqref{PDE_nonlinear} for the velocity potential that proves to be a generalisation of Kuznetsov's equation.

Via linearisation it is shown that equation \eqref{PDE_nonlinear} is also a nonlinear generalisation of the linear wave equation derived previously by \cite{SCHOLLE2020102636}. \markup{A deeper analysis of the resulting dispersion relation reveals in comparison with the classical theory (i) a reduced attenuation coefficient, (ii) a dispersion deviating stronger from a pure linear one and (iii) an upper bound for the frequency of the wave. The latter also implies a lower bound for the wavelength, which suggests a physical interpretation as a violation of the continuum hypothesis if the wavelength reaches the order of magnitude of the mean free path length of molecules. }

An obvious further generalisation of the present theory is the inclusion of thermal conduction, a promising starting point of which is the extended Lagrangian proposed by \cite{SCHOLLE2020102636}. Also a further generalisation towards non-Newtonian fluids and also solids should be possible by amending the Lagrangian with respective additional terms.


\bibliographystyle{elsarticle-harv} 
\bibliography{lit}





\end{document}